 \documentclass[
    aps,
    prl,
    reprint,
    superscriptaddress,
    nolongbibliography,
    nobibnotes,
    nofootinbib,
]{revtex4-2}

\usepackage[dvipsnames, usenames]{xcolor}
\definecolor{linkcolor}{rgb}{0.1, 0.5, 0.7}

\usepackage{acronym}
\usepackage{lmodern}
\usepackage[
	colorlinks=true,
	linkcolor=linkcolor,
	citecolor=linkcolor,
	filecolor=linkcolor,
	urlcolor=linkcolor,
	pdfusetitle,
]{hyperref}

\usepackage{amssymb}
\usepackage{amsmath}
\usepackage{physics}
\usepackage{orcidlink}
\usepackage{xspace}
\usepackage[normalem]{ulem}

\usepackage{titlesec}
\titleformat{\section}[runin]{\it}{\thesection}{0pt}{\phantomsection}[---]
\titlespacing{\section}{\parindent}{\parskip}{-\parskip}

\let\oldacknowledgments\acknowledgments
\renewcommand{\acknowledgments}{\vspace{\baselineskip}\oldacknowledgments}
\newacro{PE}[PE]{parameter estimation}
\newacro{GW}[GW]{gravitational-wave}
\newacro{GWTC-3}[GWTC-3]{third gravitational-wave transient catalog}
\newacro{GWTC-2}[GWTC-2]{second gravitational-wave transient catalog}
\newacro{GWTC-4}[GWTC-4.0]{Fourth Gravitational-Wave Transient Catalog}
\newacro{GWTC-5}[GWTC-5.0]{Fifth Gravitational-Wave Transient Catalog}
\newacro{GWTC}[GWTC]{gravitational wave transient catalog}
\newacro{LVK}[LVK]{LIGO-Virgo-Kagra collaboration}
\newacro{O4}[O4]{fourth observing run}
\newacro{O3}[O3]{third observing run}
\newacro{HMC}[HMC]{Hamiltonian Monte Carlo}
\newacro{MC}[MC]{Monte Carlo}
\newacro{NUTS}[NUTS]{No-U-Turn Sampler}
\newacro{IID}[IID]{identically and independently distributed}
\newacro{CDF}[CDF]{Cumulative Distribution Function}
\newacro{SFR}[SFR]{star formation rate}
\newacro{KDE}[KDE]{kernel density estimate}
\newacro{BBH}[BBH]{binary black-hole}
\newacro{BNS}[BNS]{binary neutron star}
\newacro{NSBH}[NSBH]{neutron star-black hole binary}
\newacro{NS}[NS]{neutron star}
\newacro{BH}[BH]{black hole}
\newacro{PISN}[PISN]{pair-instability supernova}

\newacro{PPC}[PPC]{posterior predictive check}
\newacro{MI}[MI]{mutual information}
\newacro{CAR}[CAR]{conditional autoregressive}
\newacro{ICAR}[ICAR]{intrinsic conditional autoregressive}
\newacro{PDF}[PDF]{probability density function}
\newacro{KL}[KL]{Kullback--Leibler}
\newacro{JS}[JS]{Jensen--Shannon}

\newacro{XLA}[XLA]{Accelerated Linear Algebra}
\newacro{HLV}[HLV]{Hanford, Livingston and Virgo}
\newacro{SNR}[SNR]{signal-to-noise ratio}
\newacro{IID}[IID]{independent and identically distributed}
\newacro{PSD}[PSD]{power spectral density}
\newacro{DAG}[DAG]{directed acyclic graph}

\newacro{CE}[CE]{common-envelope}
\newacro{OC}[OC]{open cluster}
\newacro{NSC}[NSC]{nuclear star cluster}
\newacro{CHE}[CHE]{chemically-homogeneous evolution}
\newacro{GC}[GC]{globular cluster}
\newacro{AGN}[AGN]{active galactic nuclei}
\newacro{SMT}[SMT]{stable mass transfer}

\usepackage{dashrule}
\usepackage[absolute,overlay]{textpos}

\newcommand{\comment}[1]{}

\newcommand{\chieff}{\ensuremath{\chi_\mathrm{eff}}\xspace}
\newcommand{\chip}{\ensuremath{\chi_\mathrm{p}}\xspace}

\newcommand{\Msun}{\ensuremath{\mathrm{M}_\odot}\xspace}

\newcommand{\skewnormal}{\textsc{Skewnormal}\xspace}
\newcommand{\bivskewnormal}{\textsc{Bivariate Skewnormal}\xspace}
\newcommand{\gtwou}{\textsc{Gaussian+Two-Uniforms}\xspace}
\newcommand{\PixelPop}{\textsc{PixelPop}\xspace}
\newcommand{\Mixture}{\textsc{Mixture}\xspace}
\newcommand{\supmat}{Supplementary Material\xspace}

\newcommand{\ligo}{\affiliation{LIGO Laboratory, Massachusetts Institute of Technology, Cambridge, MA 02139, USA}}
\newcommand{\mki}{\affiliation{Kavli Institute for Astrophysics and Space Research, Massachusetts Institute of Technology, Cambridge, MA 02139, USA}}
\renewcommand{\mit}{\affiliation{Department of Physics, Massachusetts Institute of Technology, Cambridge, MA 02139, USA}}

\begin{document}
\begin{textblock*}{4cm}(0.5pt,5pt)
    \includegraphics[width=4cm]{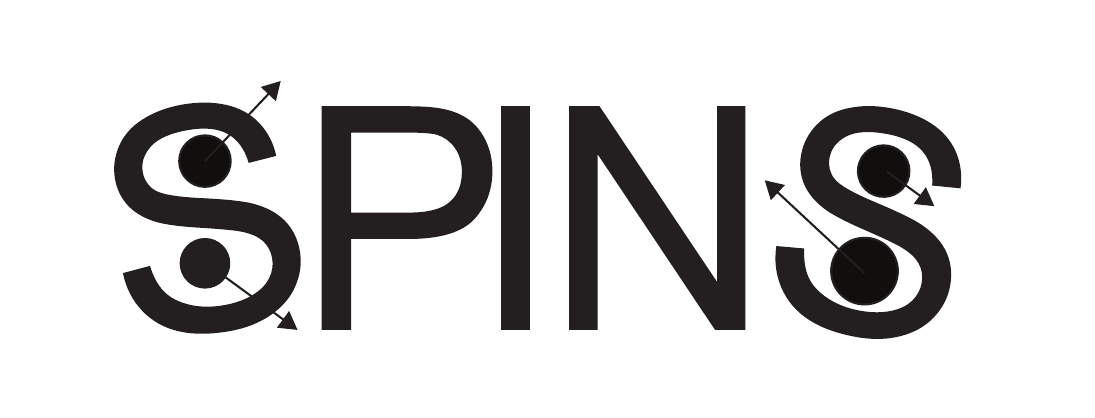}
\end{textblock*}

\title{Evidence for additional structure in the effective spin distribution hints at multiple formation pathways in GWTC-5.0}

\author{Sofía Álvarez-López\,\orcidlink{0009-0003-8040-4936}}
\email{sofiaal@mit.edu}
\ligo\mki\mit

\author{Jack Heinzel\,\orcidlink{0000-0002-5794-821X}}
\ligo\mki\mit

\author{Salvatore Vitale\,\orcidlink{0000-0003-2700-0767}}
\ligo\mki\mit
\collaboration{Society of Physicists Interested in Non-aligned Spins, SPINS}
\thanks{\url{www.sites.mit.edu/spins}}
\date{\today}

\begin{abstract}
The distribution of the effective inspiral spin (\chieff) of the binary black holes detected by LIGO-Virgo-KAGRA can shed light on their formation pathways. We analyze the GWTC-5.0 dataset with two models---one flexible, one fully parametric---that jointly describe \chieff and primary mass. We clarify that the previously-reported skewness in the \chieff distribution is better understood as additional structure beyond a non-skewed Gaussian bulk centered at small \chieff. This additional structure extends to larger $|\chieff|$, a result previously reported using GWTC-4.0 data. We measure the asymmetry of the distribution of \chieff outside the Gaussian bulk from the data. With both the parametric and the flexible analyses, we find suggestive evidence for a mass-dependent excess of positive \chieff over negative ones outside the Gaussian bulk. {We find only a mild statistical preference for a negative \chieff component outside the Gaussian bulk. This preference is highest at $m_1 \in [47,65]\,\Msun$, with $13\text{:}1$ odds. Excluding the event GW191109, whose data have known data-quality issues, reduces these odds to $5\text{:}1$.} If \chieff outside the Gaussian bulk are produced by hierarchical mergers---as has been suggested---then a fraction of those mergers may be produced in environments that can generate a surplus of binaries with positive \chieff, such as the disks of active galactic nuclei. 
\end{abstract}

\maketitle

\section{Introduction}

With the recent release of the \acf{GWTC-5}~\citep{LIGOScientific:2026sit, LIGOScientific:2026ifv,LIGOScientific:2026wfs} by the LIGO~\citep{TheLIGOScientific:2014jea}, Virgo~\citep{TheVirgo:2014hva}, and KAGRA~\citep{KAGRA:2020tym} collaborations (\acs{LVK}), \ac{GW} observations of merging \acfp{BBH} now include hundreds of events, enabling increasingly detailed studies of the astrophysical processes that shape the population~\citep{LIGOScientific:2026ctl}. \acp{BBH} are predicted to form through a diverse range of astrophysical channels~\citep{Mapelli:2020vfa,Ivanova:2012vx,Bavera:2020uch,Gerosa:2021mno,Mapelli:2019bnp,Mandel:2021smh}, which leave distinct, albeit potentially overlapping, imprints on the observed distributions of masses, spins, and redshifts~\citep{Mapelli:2021taw,Zevin:2020gbd,Mandel:2018hfr,Colloms:2025hib,Biscoveanu:2026ikx}.

Disentangling the formation channels that contribute to the \ac{BBH} population requires inferring parameters that are both well-measured in \ac{GW} data and informative about the astrophysical origin of the binaries. One such parameter is the effective inspiral spin, $\chieff$, the mass-weighted projection of the component \ac{BH} spins along the binary orbital angular momentum~\citep{Racine:2008qv,Gerosa:2017kvu}. Although $\chieff$ is among the best-measured spin observables~\citep{Vitale:2016avz}, similar $\chieff$ signatures can arise from distinct astrophysical channels~\citep{Farr:2017uvj,Zevin:2020gbd,Bouffanais:2021wcr,Payne:2024ywe,Hussain:2026pfm}. 

Isolated binary evolution is expected to produce a narrow \chieff distribution centered near zero~\citep{Gerosa:2017kvu,Farr:2017uvj,Belczynski:2017gds,Fuller:2019sxi}. However, uncertainties in stellar angular-momentum transport, binary interactions, and tidal spin-up~\citep{Qin:2018vaa,Fuller:2022ysb,Bavera:2020inc,Olejak:2021iux,Ma:2023nrf,Olejak:2024qxr}, as well as pathways such as chemically homogeneous evolution~\citep{deMink:2009jq,Mandel:2015qlu, deMink:2016vkw, Marchant:2016wow,duBuisson:2020asn,Marchant:2023ncp}, may result in some isolated binaries having large, positive \chieff. 
\acp{BBH} can also form through dynamical channels. In dense stellar environments, first-generation binaries, whose constituents form directly by stellar collapse, are expected to have a narrow \chieff distribution centered near zero. Hierarchical mergers, which involve at least one remnant of a previous \ac{BBH} coalescence, can have high spin magnitudes and nearly isotropic orientations, resulting in a broad \chieff distribution symmetric around zero~\citep{OLeary:2005vqo,Pretorius:2005gq,Buonanno:2007sv,Tichy:2008du,Rezzolla:2007rz,Hofmann:2016yih}. In \ac{AGN} disks, gas accretion and torques can spin up \acp{BH} and align their spins with the disk angular momentum, producing a high-spin population with preferentially positive \chieff~\citep{Antonini:2016gqe,Mckernan:2017ssq,McKernan:2019beu,Stone:2016wzz,Tagawa:2019osr,Tagawa:2021ofj,Bartos_2017,Yang:2020lhq,McKernan:2024kpr,Cook:2024ajp,Delfavero:2024fcc} (although negative-$\chieff$ mergers may also occur~\citep{Fabj:2025vza}). Therefore, while multiple channels may produce aligned-spin binaries ($\chieff>0$), anti-aligned ones ($\chieff<0$) are difficult to form in environments other than dense stellar clusters. 

The two most recent \acs{LVK} population analyses used a skewed Gaussian model for the \chieff distribution, finding evidence that---while centered at or close to $0$---it \emph{must} be positively skewed~\citep{LIGOScientific:2025pvj,LIGOScientific:2026ctl}, suggesting an excess of aligned-spin systems~\citep{Banagiri:2025dxo}. Meanwhile, their flexible \PixelPop model~\citep{Heinzel:2024jlc} is consistent with zero skewness in \ac{GWTC-5} data~\citep{LIGOScientific:2026ctl}. Other studies~\citep{Antonini:2024het,Antonini:2025zzw,Tong:2025xir,Plunkett:2026pxt} model $\chieff$ as a mass-dependent mixture of a slowly spinning Gaussian bulk containing most of the sources and a broad, approximately uniform component motivated by hierarchical mergers in clusters. In these models, the broad component preferentially contributes in two regimes: at $m_1 \in [13,20]\,\Msun$, and above $m_1 \approx 45\,\Msun$, where hierarchical assembly is a natural candidate to populate the \ac{PISN} gap~\citep{Fowler:1964zz,Barkat:1967zz,1967ApJ...148..803R,Heger:2002by,Farmer:2019,Stevenson:2019rcw,Farmer:2020xne,Woosley:2021xba}. Additional population analyses have also inferred aligned, high-spin subpopulations at high masses and interpreted them as possible signatures of assembly in \ac{AGN} disks~\citep{Li:2025rhu,Li:2025iux,Wang:2025nhf,Bartos:2026xlt,Padhyegurjar:2026scg}. These approaches point to distinct formation pathways but rely on strong assumptions about how spin structure is modeled and interpreted. 

In this \emph{Letter}, we use both flexible and parametric models and find no evidence that the Gaussian bulk of the \chieff distribution is skewed. We confirm the mass-dependent \chieff structure reported by others, but suggest that this structure may be asymmetric: we find tentative evidence for an excess of positive over negative \chieff outside the bulk. A negative \chieff component is only weakly favored, and only at high masses. If confirmed with larger datasets, these findings would call into question the interpretation of all \chieff outside of the Gaussian bulk as hierarchical mergers in clusters, pointing instead to contributions from multiple formation channels.

\section{Results}~\label{sec:results} Following Ref.~\citep{LIGOScientific:2026ctl}, we analyze the 259 \acp{BBH} from \ac{GWTC-5} with false-alarm rates below $1\,\mathrm{yr}^{-1}$. We employ the standard hierarchical Bayesian framework for gravitational-wave population inference~\citep{Mandel:2018mve,Thrane:2018qnx,Vitale:2020aaz,Essick:2023upv}. 

\textbf{Evidence for additional structure in the effective spin distribution:} We first analyze the population with a flexible \Mixture model~\citep{Alvarez-Lopez:inprep} that combines a parametric component with a nonparametric \PixelPop component (c.f. Ref.~\citep{Godfrey:2026pbc}). We write the joint distribution of $(m_1,\chieff)$ as 
\begin{equation}\label{eq:mixture-def}
\begin{aligned}
p_\mathrm{mix}(m_1,\chieff)
&= \xi\, p_\mathrm{PP}(m_1,\chieff) \\
&\quad + (1-\xi)\, p_\mathrm{par}(m_1,\chieff),
\end{aligned}
\end{equation}
where $p_\mathrm{par}$ is the parametric component, $p_\mathrm{PP}$ is the nonparametric \PixelPop model~\citep{Heinzel:2024jlc,Heinzel:2024hva,Alvarez-Lopez:2025ltt}, and $\xi \in [0,1]$ controls their relative contribution.  The parametric component is built from a mixture of two Gaussian peaks with a broken power law tapered at low masses for primary mass~\citep{LIGOScientific:2025pvj,LIGOScientific:2026ctl}, and a skewed normal distribution truncated to $[-1,1]$ for effective spin~\citep{Banagiri:2025dxo,LIGOScientific:2025pvj}. When referring to the parametric sector of this model, we will call it \skewnormal model. Both components share the same mass ratio $q$, redshift $z$, and effective precessing spin \chip distributions, see the \supmat.

\begin{figure}
\centering
\includegraphics[width=\columnwidth]{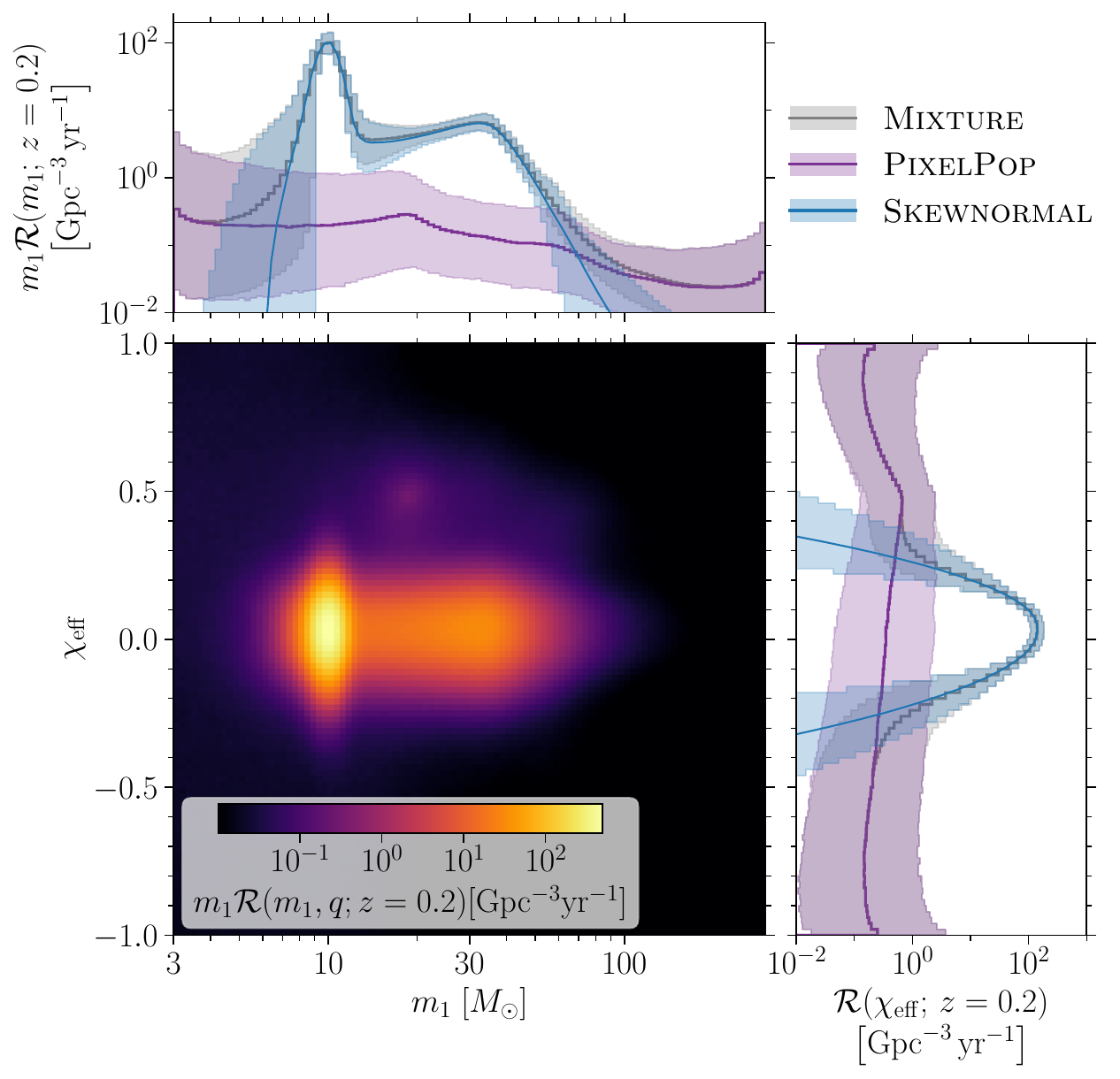}
\caption{Inferred comoving merger-rate density $\mathcal{R}(m_1,\chieff;z=0.2)$. The central panel shows the median of the two-dimensional merger-rate posterior; brighter regions indicate a higher inferred merger rate. The upper (right-hand) panel shows the rate marginalized over the source parameter on the vertical (horizontal) axis. The solid lines and shaded bands represent the posterior median and central 90\% credible region of the full mixture model (gray), the parametric component (blue), and the \PixelPop component (purple).
}
\label{fig:mixture-corner}
\end{figure}

Figure~\ref{fig:mixture-corner} shows the inferred joint and marginal comoving merger rate density. Outside regions of low detectability, such as~$m_1 \leq 5\,\Msun$ (see discussion in Sec.~3 of Ref.~\citep{Alvarez-Lopez:2025ltt}), the \PixelPop component recovers structure beyond the parametric model, most prominently around $\chieff\approx0.5$ and $m_1\in [15,25]\,\Msun$, overlapping with a previously-reported low-significance peak in the primary mass distribution near $\sim20\,\Msun$~\citep{Tiwari:2020otp,Tiwari:2023xff,Edelman:2021zkw,Toubiana:2023egi,Galaudage:2026opk}. Consistently, the posterior on the mixing fraction is measured away from zero at $\xi = 2.7_{-2.1}^{+7.9}\%$ (90\% symmetric credible interval), indicating that the data favor a correction to the parametric model. We also find support at $\chieff\approx0.5$ (albeit at lower merger rate) out to $m_1\approx100\,\Msun$. These additional features are robust to the removal of GW241011, a $\chieff \sim 0.5$, unequal-mass-ratio event with possible hierarchical origin~\citep{LIGOScientific:2025brd} (see \supmat).

\begin{figure*}
\centering
\includegraphics[width=\linewidth]{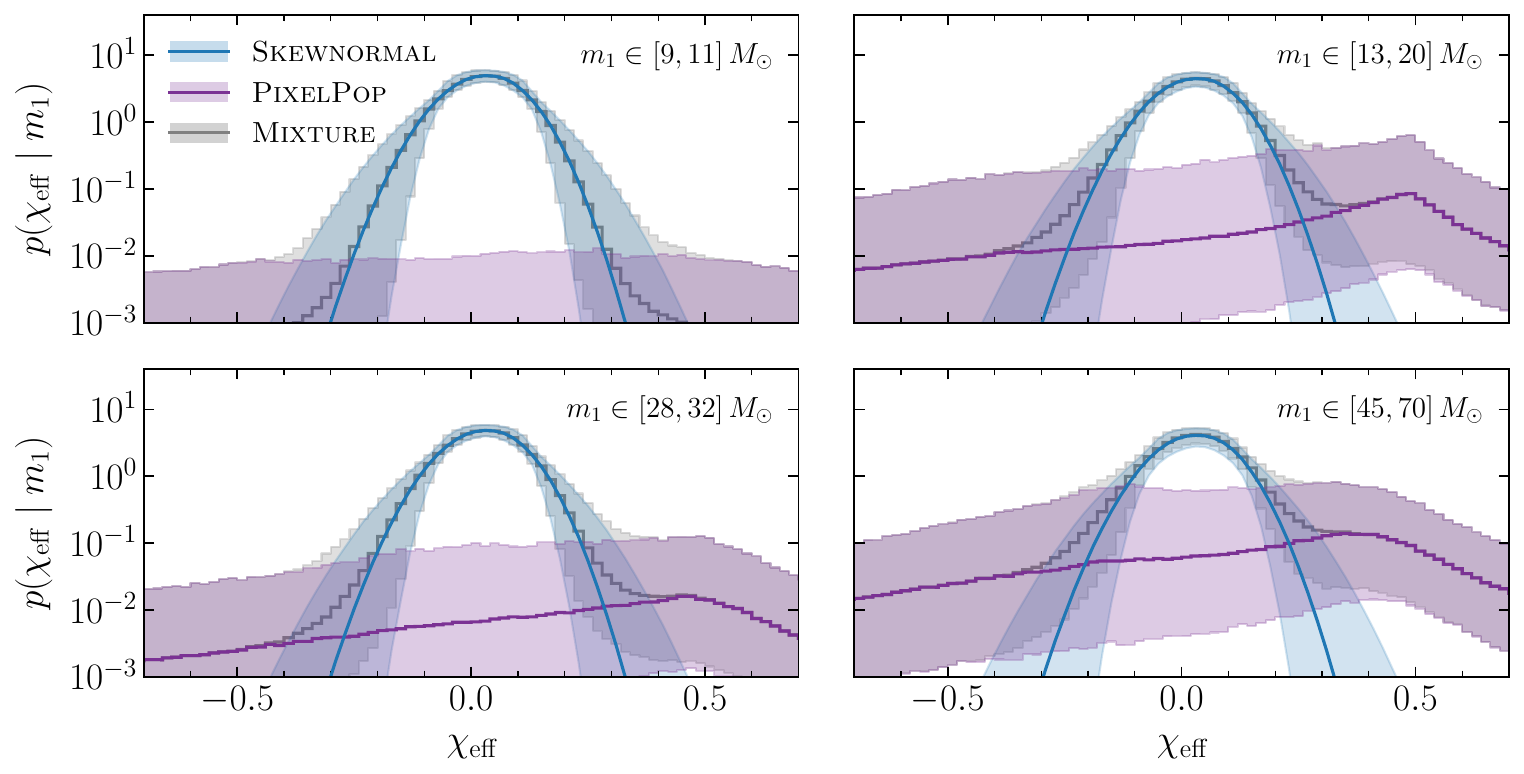}
\caption{Conditional \chieff distributions $p_{\mathrm{mix}}(\chieff \mid m_1)$ for $m_1 \in [9,11]\,\Msun$ (top left), $m_1 \in [13,20]\,\Msun$ (top right), $m_1 \in [28,32]\,\Msun$ (bottom left), and $m_1 \in [45,70]\,\Msun$ (bottom right). The solid lines and shaded bands show the posterior median and central 90\% credible region of the full \Mixture model (light gray), together with its parametric (blue) and \PixelPop (purple) components. 
}
\label{fig:slices}
\end{figure*}

Furthermore, the skewness parameter of the \skewnormal component is consistent with zero, $\epsilon_{\chi_\mathrm{eff}} = -0.07^{+0.56}_{-0.44}$. This is in contrast to Refs.~\citep{LIGOScientific:2025pvj,LIGOScientific:2026ctl}, who used a single skewed Gaussian and found positive skewness at $\geq99.3\%$ credibility. While our model could have suppressed the \PixelPop component, yielding a single skewed Gaussian, it prefers a solution with a non-skewed Gaussian plus an additional component.

We use the (natural) log Bayes factor $\ln \mathcal{B}_{\textsc{PP}/\mathrm{SN}}$ to quantify the odds that any given event belongs to the \PixelPop component rather than the \skewnormal component; see the \supmat for details. Ranking events by this quantity, the highest-ranked event is GW241011, with $\ln \mathcal{B}_{\textsc{PP}/\mathrm{SN}} = 4.1$ (a 98.3\% probability of belonging to the \PixelPop component). Among the 10 highest-ranked events (see Tab.~\ref{tab:individual-events} in the \supmat) 9 have $\chieff>0$ at 90\% credibility. However, these 10 events don't share any other common characteristics (see Fig.~\ref{fig:event-corner-plots}): their primary masses span a wide range $m_1 \in[13,165]\,\Msun$ and their mass ratios do not cluster around any particular value (e.g., $q\lesssim 0.6$ as would be expected from hierarchical mergers~\citep{Gerosa:2017kvu,Rodriguez:2019huv,Kimball:2020opk,Vijaykumar:2026zjy}).

We note that the residual structure inferred by the \Mixture model need not be symmetric in $\chieff$, as would be expected from hierarchical mergers in dense stellar clusters~\citep{Baibhav:2020xdf,Antonini:2024het}. Integrating over $m_1\in[3,200]\,\Msun$, we find that aligned-spin systems outnumber anti-aligned ones with 75\% probability (excluding the tails of the distribution, $|\chieff| \gtrsim 0.8 $, where we don't expect $\geq\mathcal{O}(1)$ detections). While the \PixelPop component remains consistent with symmetry, and even allows for an anti-aligned spin excess, the mild preference for aligned spins motivates examining whether the positive-$\chieff$ excess is localized to particular mass ranges. 

Figure~\ref{fig:slices} shows the conditional distribution $p_\mathrm{mix}(\chieff \mid m_1)$, and the individual components of the \Mixture model, for $m_1 \in \{[9, 11], [13,20], [28,32], [45,70]\}\,\Msun$. We choose mass ranges around $10\,\Msun$ and $30\,\Msun$ because they correspond approximately to the known features of the primary-mass distribution~\cite{LIGOScientific:2025pvj,KAGRA:2021duu,LIGOScientific:2026ctl}, while $[13,20]\,\Msun$ and $[45,70]\,\Msun$ fall within regions where previous analyses have identified possible hierarchical-merger contributions~\citep{Tong:2025xir,Plunkett:2026pxt,Antonini:2025ilj,Antonini:2025zzw,LIGOScientific:2026ctl}.

At $m_1 \in [9,11]\,\Msun$, the distribution is dominated by the \skewnormal component, centered at $\chieff=0.04_{-0.06}^{+0.05}$, with little support for residual $\chieff$ structure. This behavior is consistent with previous population analyses that identified the low-mass peak as compatible with isolated binary evolution~\citep{Godfrey:2023oxb,KAGRA:2021duu,LIGOScientific:2026ctl,Cheng:2026bpc,Galaudage:2026opk}, which is expected to form slowly spinning \acp{BH} with mildly positive $\chieff$~\citep{Dominik:2014yma, Belczynski:2017gds,Giacobbo:2018etu,Wiktorowicz:2019dil,Neijssel:2019irh,vanSon:2022myr,vanSon:2022ylf,KAGRA:2021duu,LIGOScientific:2026ctl}. At $m_1 \in [13,20]\,\Msun$ the \PixelPop component develops support near $\chieff \simeq 0.5$, with an 82\% probability of an aligned-spin excess. Near the $\sim~30\,\Msun$ feature, the \PixelPop distribution shows a mild positive shoulder, although at lower probability density than in the $m_1\in [13,20]\,\Msun$ interval, with an 80\% probability of positive excess. At higher masses, $m_1 \in [45,70]\,\Msun$, the \PixelPop component again peaks near $\chieff \simeq 0.5$, but increased support at negative $\chieff$ points to a non-negligible anti-aligned contribution. The probability of an excess of positive-$\chieff$ systems reduces to 77\% in this mass range.

The \Mixture model suggests that, outside of the Gaussian bulk, the positive and negative sides of the $\chieff$ distribution may evolve differently with primary mass. In particular, in regions that previous analyses have identified as possibly populated by hierarchical mergers in dense stellar clusters~\citep{Antonini:2024het,Antonini:2025zzw,Tong:2025xir,Plunkett:2026pxt}, the \PixelPop component mildly favors a positive-$\chieff$ excess. Such an excess would not be expected if all those \acp{BBH} were indeed hierarchical cluster binaries.

\textbf{Is there an excess of aligned, high-spin \acp{BBH}?} The flexibility of our mixture model comes at the expense of large statistical uncertainties. Therefore, having used it to illuminate the global morphology of the posterior probability, we now follow up with a parametric correlated \chieff--$m_1$ model, dubbed \gtwou, to quantify how the relative contributions of aligned and anti-aligned systems depend on primary mass. 

Our model is similar in form to Refs.~\citep{Antonini:2024het,Antonini:2025zzw,Tong:2025xir}, but with two key differences (see the \supmat). First, rather than mixing a Gaussian with a single uniform component symmetric about $\chieff=0$, we employ a \skewnormal component (to capture the slowly spinning bulk of the population, allowing for a potential skewness) and two half-uniform components on either side of $\chieff=0$. We infer the outer edges of the positive and negative half-uniform components from the data. Second, each component has its own mixing fraction that varies independently with primary mass (see Table~\ref{tab:mixing-fractions} for a physical interpretation). We model the mass dependence of each mixing fraction with a cubic spline, rather than assuming a priori knowledge on the existence or location of transitions. 

\begin{table}[t]
\begin{tabular}{l  p{0.75\columnwidth}}
\hline\hline
\textbf{Symbol} & \textbf{Interpretation} \\ \hline
$\xi_G(m_1)$ & Fraction of the population in the slowly spinning Gaussian bulk. \\
$\xi_U^+(m_1)$ & Within the non-Gaussian component, fraction of systems with $\chieff>0$ (i.e., aligned). \\
\hline\hline
\end{tabular}
\caption{Symbols used and interpretation of the mixing fraction parameters for the \gtwou model.}\label{tab:mixing-fractions}
\end{table}

\begin{figure}
\centering
\includegraphics[width=\columnwidth]{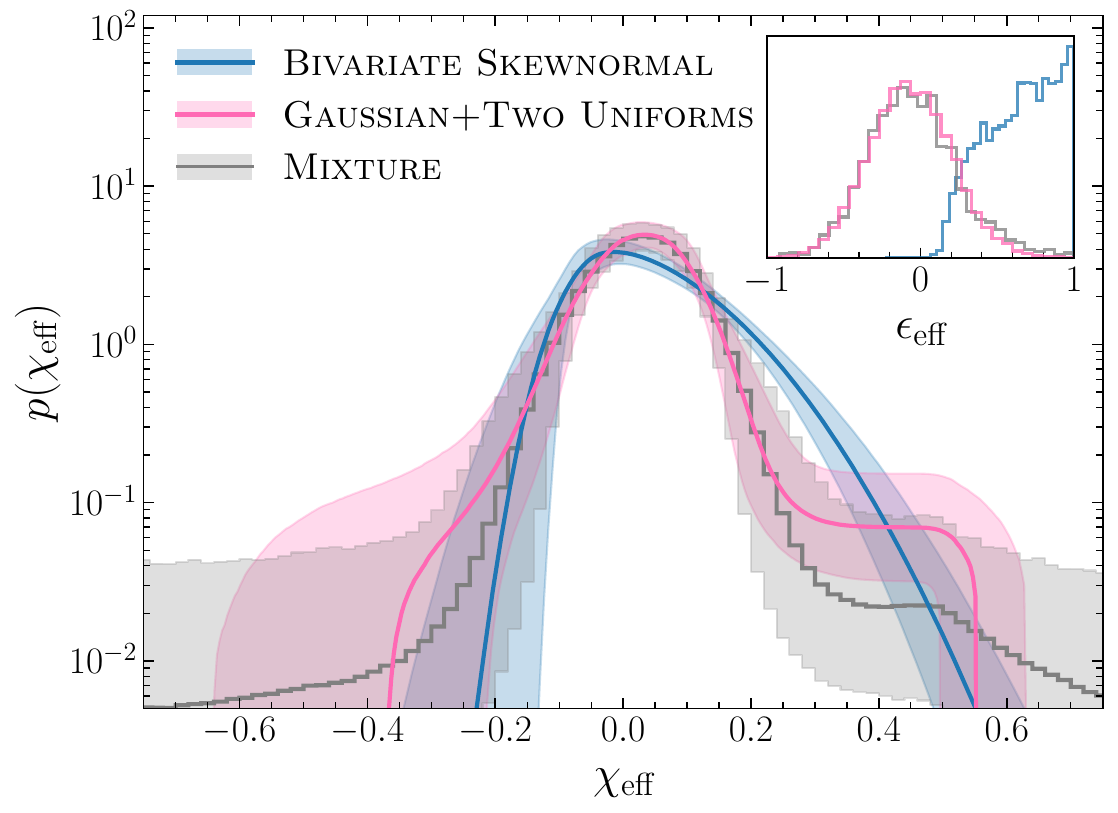}
\caption{Posterior distributions of the effective spin, $p(\chieff)$, inferred using the \gtwou model (pink) and the flexible \Mixture model (gray), with the \bivskewnormal\ fit from Ref.~\citep{LIGOScientific:2026ctl} shown for comparison (blue). The inset shows the posterior on the skewness of the Gaussian component of each model, $\epsilon_\mathrm{eff}$.
}
\label{fig:marginal-chieff}
\end{figure}

Figure~\ref{fig:marginal-chieff} shows the $\chieff$ distribution marginalized over $m_1$ for the \gtwou and \Mixture models, together with the \bivskewnormal \acs{LVK} model (marginalized over \chip) for comparison~\citep{ligo_scientific_collaboration_2026_20292639}. Both \gtwou and \Mixture are in tension with the \acs{LVK} \bivskewnormal distribution at 90\% credibility over the range $\chieff \in [0.2, 0.4]$. The \gtwou model is also strongly favored over the \bivskewnormal baseline, with natural log Bayes factor $\ln \mathcal{B} = 12$\footnote{To calculate this Bayes factor, we implement the \bivskewnormal model in \textsc{GWPopulation}~\citep{Talbot:2024yqw}, since the \ac{GWTC-5} data release does not report evidences for this model.}. As in the \Mixture model, the data are consistent with a non-skewed ($\epsilon_{\chieff} = -0.07_{-0.43}^{+0.45}$) slowly spinning Gaussian bulk. We infer $\chieff^{\max}=0.55^{+0.08}_{-0.06}$ for the right edge of the positive uniform component, above the value $\chieff=0.47$ one would expect from hierarchical mergers at 99\% credibility. In contrast, the left edge of the negative half-uniform is more weakly constrained at $\chieff^{\min}=-0.36^{+0.19}_{-0.29}$. 

\begin{figure}
\centering
\includegraphics[width=\columnwidth]{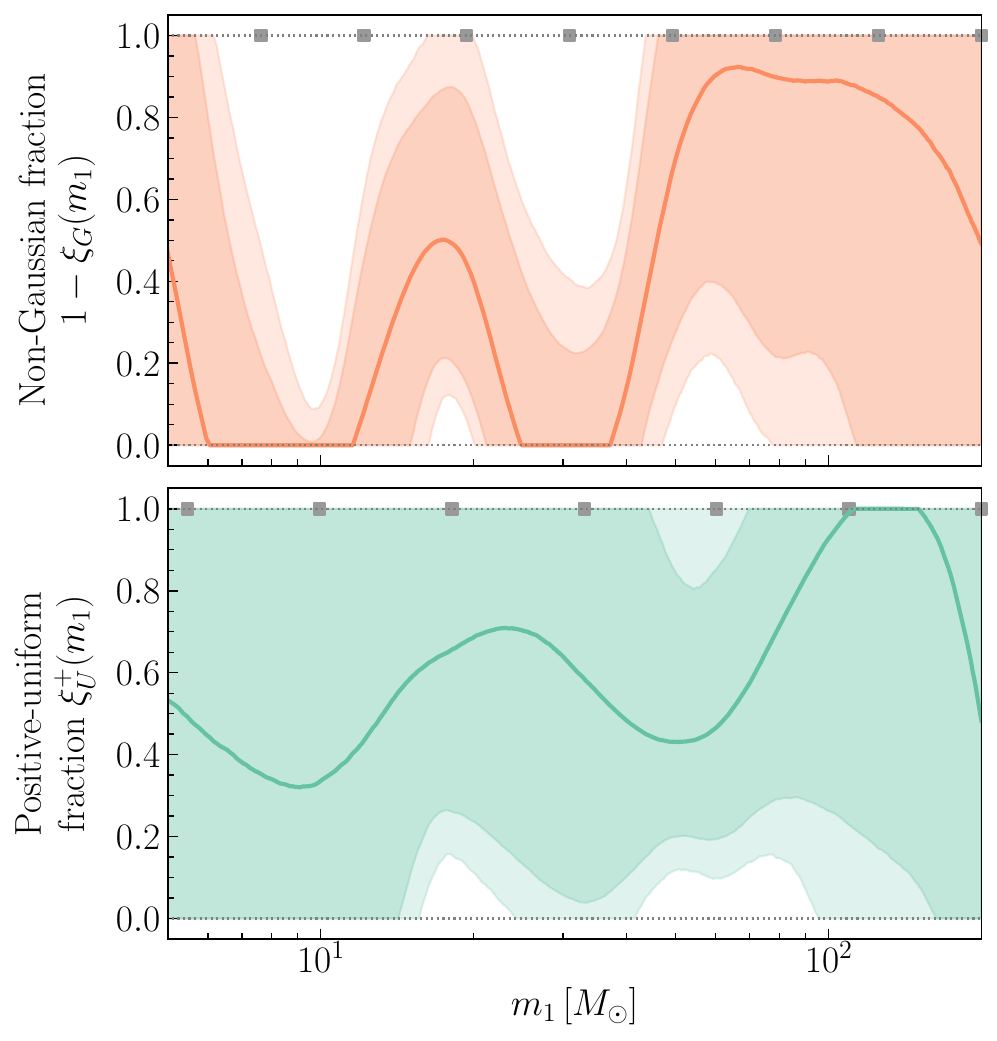}
\caption{Mixing fractions inferred with the \gtwou model as a function of primary mass. Solid curves show the posterior median, shaded regions the 90\% and 99\% credible intervals. Gray squares mark the spline node locations. \textit{Top:} $1-\xi_G(m_1)$, the fraction of the population outside the slowly spinning Gaussian bulk. \textit{Bottom:} $\xi_U^+(m_1)$ the fraction of binaries outside the Gaussian bulk that have $\chieff>0$.
}
\label{fig:mixing-fractions}
\end{figure}

The mass-dependent mixing fractions of the \gtwou model (Tab.~\ref{tab:mixing-fractions}) can be used to quantify where the data are consistent with a Gaussian bulk alone, and where additional $\chieff$ contributions---aligned or anti-aligned---are required. The top panel of Fig.~\ref{fig:mixing-fractions} shows $1-\xi_G(m_1)$, the fraction of the population \emph{outside} the slowly spinning Gaussian bulk. The bottom panel shows $\xi_U^+(m_1)$, the fraction of the non-Gaussian component with $\chieff>0$.

Below $m_1 = 200\,\Msun$, the median of the non-Gaussian fraction is consistent with zero in two mass ranges: $m_1 \in [6, 11]\,\Msun$ and $m_1 \in [25, 37]\,\Msun$. Near $m_1 \simeq 10\,\Msun$, we measure $1-\xi_G(m_1) < 0.09$ at 99\% credibility, indicating that no additional contribution is required, in agreement with the \Mixture model. At $m_1 \simeq 30\,\Msun$, the non-Gaussian fraction is again consistent with zero, although with broader uncertainties, $1-\xi_G(m_1) < 0.42$ at 99\% credibility. If a non-Gaussian contribution is present, it is preferentially associated with the positive uniform: we measure $\xi_U^+(m_1) > 0$ at 97\% credibility, consistent with the mild positive shoulder at $m_1 \in [28,32]\,\Msun$ in the \Mixture model (see bottom-left panel in Fig.~\ref{fig:slices}). 

At $m_1 \in [16, 20]\,\Msun$ and {$m_1 \in [47, 78]\,\Msun$, the non-Gaussian fraction, $1-\xi_G(m_1)$, is constrained away from zero at 99\% credibility. These mass ranges are consistent with previous analyses that identified possible hierarchical-merger contributions~\citep{Tong:2025xir,Plunkett:2026pxt,Antonini:2025ilj,Antonini:2025zzw,LIGOScientific:2026ctl}. However, a nonzero value of $1-\xi_G(m_1)$ only indicates that structure beyond the Gaussian bulk is required; it does not imply that this structure is symmetric in $\chieff$. 

In the mass range $m_1 \in [16,20]\,\Msun$, the positive uniform is required with $\ln\mathcal{B}=8.4 \pm 1.0$; see the \supmat for details of the Bayes-factor calculation. In contrast, the negative uniform is only mildly favored ($\ln\mathcal{B}=0.77 \pm 0.03$), indicating that the non-Gaussian contribution could be dominated, or even entirely explained, by systems with positive $\chieff$. Although we cannot exclude that both components contribute, the positive uniform is more strongly required than the negative one, consistent with the \Mixture model results (see top-right panel of Fig.~\ref{fig:slices}). 

The picture changes at higher masses. For $m_1 \in [47,65]\,\Msun$, we still require the positive uniform component with a Bayes factor $\ln B = 6.5 \pm 0.5$. However, in contrast to the lower-mass range, we now find $\xi_U^+(m_1) < 1$ at 95\% credibility ($\ln\mathcal{B} = 2.6 \pm 0.1$ for the presence of the negative uniform component), suggesting a contribution of anti-aligned-spin binaries to the non-Gaussian part of the population. This is the region of the mass spectrum where our analysis most favors the presence of negative-\chieff sources expected in the context of hierarchical mergers in dense clusters. However, the preference for a negative component depends strongly on GW191109, an event whose \chieff posterior is affected by known data-quality issues~\citep{Udall:2024ovp}. In a leave-one-out analysis excluding this event (see the \supmat), the natural-log Bayes factor for the presence of a negative uniform component decreases to a mild $\ln\mathcal{B}=1.6\pm0.1$}.

\section{Discussion and conclusion}
In this \emph{Letter}, we presented strong evidence that the \chieff distribution of merging \acp{BBH} includes mass-dependent structure beyond a slowly-spinning Gaussian bulk. Using two different models---neither of which prescribes if and how \chieff varies with mass---we identified two mass ranges where the residual structure in $\chieff$ is most prominent. 
At $m_1 \in [16,20]\,\Msun$, we found an excess of aligned-spin systems (i.e., with $\chieff>0$); the absence of an anti-aligned contribution ($\chieff<0$) is only mildly disfavored. 

Although hierarchical mergers in dense stellar clusters---possibly including higher-generation mergers---can explain a broad and symmetric \chieff distribution~\citep{OLeary:2005vqo,Pretorius:2005gq,Buonanno:2007sv,Tichy:2008du,Rezzolla:2007rz,Hofmann:2016yih,Baibhav:2020xdf,Antonini:2024het,Farah:2026jlc,Franciolini:2022iaa} they can't easily explain an excess of positive \chieff.
Formation pathways that preferentially produce aligned spins, such as tidally spun-up isolated binaries~\citep{Qin:2018vaa,Fuller:2022ysb,Bavera:2020inc,Olejak:2021iux,Ma:2023nrf,Olejak:2024qxr}, chemically homogeneous evolution~\citep{deMink:2009jq,Mandel:2015qlu, deMink:2016vkw, Marchant:2016wow,duBuisson:2020asn,Marchant:2023ncp}, or \acp{BBH} in \ac{AGN} disks~\citep{Antonini:2016gqe,Mckernan:2017ssq,McKernan:2019beu,Stone:2016wzz,Tagawa:2019osr,Tagawa:2021ofj,Bartos_2017,Yang:2020lhq,McKernan:2024kpr,Cook:2024ajp,Delfavero:2024fcc,Li:2025rhu,Li:2025iux,Wang:2025nhf,Bartos:2026xlt,Padhyegurjar:2026scg}, may contribute alongside (or instead of) mergers in clusters. Around $\sim 30\,\Msun$, we found that no additional $\chieff$ structure beyond the slowly spinning Gaussian bulk is \textit{required}. If present, however, our models indicate that it lies primarily on the positive-$\chieff$ side, where any of the channels producing preferentially aligned spins could contribute. 

We only find evidence for a negative \chieff component at $m_1 \in [47,65]\,\Msun$, although the evidence is marginal and largely driven by GW191109, a GW source with known data-quality issues~\citep{Udall:2024ovp}. 
At these masses, \ac{PISN} is expected to prevent \ac{BH} formation through standard stellar collapse~\citep{Fowler:1964zz,Barkat:1967zz,1967ApJ...148..803R,Heger:2002by,Farmer:2019,Stevenson:2019rcw,Farmer:2020xne,
Woosley:2021xba}, making \acp{BH} in this regime natural candidates for hierarchical assembly~\citep{Rodriguez:2019huv,Gerosa:2021mno,Baibhav:2020xdf, Torniamenti:2024uxl}. Our result is consistent with other studies identifying subpopulation of sources that could be explained as hierarchical-mergers in this region~\citep{Antonini:2024het,Antonini:2025zzw,Tong:2025xir,
Plunkett:2026pxt,Antonini:2025ilj,Galaudage:2026opk,Ray:2026uur,Pierra:2024fbl}, though see Refs.~\cite{Banagiri:2025dmy,Berti:2025usa,Hussain:2026pfm}. However, we also found a 77\% probability that spin-aligned systems outnumber anti-aligned ones, suggesting additional contributions from channels that preferentially produce aligned spins. Besides hierarchical assembly, some isolated-binary pathways can also populate the \ac{PISN} gap~\citep[e.g.,][]{Popa:2025dpz}.  

Taken together, our results suggest that the sources outside the slowly-spinning \chieff Gaussian bulk might not originate from a single formation channel. In particular, that not all these binaries may be formed as hierarchical mergers in dense stellar clusters. We also offered a cautionary tale about strong parametric models, showing they can produce measurements in tension with more agnostic frameworks. Larger \ac{GW} catalogs will help disentangle the formation channels that shape the \ac{BBH} population, provided analysts use models that capture the rich complexity of the data.

\begin{acknowledgments}
We thank Matthew Mould, Cailin Plunkett, Noah Wolfe, Shanika Galaudage, Thomas Dent, Fabio Antonini, and Sylvia Biscoveanu for useful discussions. We thank Sharan Banagiri for serving as our internal LVK reviewer.
S.A.-L., J.~H. and S.~V. are partially supported by the NSF grant No. PHY-2045740.
The authors are grateful for computational resources provided by
the LIGO Laboratory supported by National Science Foundation Grants PHY-0757058 and PHY-0823459.
The hyperposterior samples produced for our analyses will be made available at the time of publication.
This material is based upon work supported by NSF's LIGO Laboratory which is a major facility fully funded by the National Science Foundation and has made use of data or software obtained from the Gravitational Wave Open Science Center (gwosc.org), a service of the LIGO Scientific Collaboration, the Virgo Collaboration, and KAGRA.
\end{acknowledgments}

\bibliography{draft}

\newacro{PE}[PE]{parameter estimation}
\newacro{GW}[GW]{gravitational-wave}
\newacro{GWTC-3}[GWTC-3]{third gravitational-wave transient catalog}
\newacro{GWTC-2}[GWTC-2]{second gravitational-wave transient catalog}
\newacro{GWTC-4}[GWTC-4.0]{Fourth Gravitational-Wave Transient Catalog}
\newacro{GWTC-5}[GWTC-5.0]{Fifth Gravitational-Wave Transient Catalog}
\newacro{GWTC}[GWTC]{gravitational wave transient catalog}
\newacro{LVK}[LVK]{LIGO-Virgo-Kagra collaboration}
\newacro{O4}[O4]{fourth observing run}
\newacro{O3}[O3]{third observing run}
\newacro{HMC}[HMC]{Hamiltonian Monte Carlo}
\newacro{MC}[MC]{Monte Carlo}
\newacro{NUTS}[NUTS]{No-U-Turn Sampler}
\newacro{IID}[IID]{identically and independently distributed}
\newacro{CDF}[CDF]{Cumulative Distribution Function}
\newacro{SFR}[SFR]{star formation rate}
\newacro{KDE}[KDE]{kernel density estimate}
\newacro{BBH}[BBH]{binary black-hole}
\newacro{BNS}[BNS]{binary neutron star}
\newacro{NSBH}[NSBH]{neutron star-black hole binary}
\newacro{NS}[NS]{neutron star}
\newacro{BH}[BH]{black hole}
\newacro{PISN}[PISN]{pair-instability supernova}

\newacro{PPC}[PPC]{posterior predictive check}
\newacro{MI}[MI]{mutual information}
\newacro{CAR}[CAR]{conditional autoregressive}
\newacro{ICAR}[ICAR]{intrinsic conditional autoregressive}
\newacro{PDF}[PDF]{probability density function}
\newacro{KL}[KL]{Kullback--Leibler}
\newacro{JS}[JS]{Jensen--Shannon}

\newacro{XLA}[XLA]{Accelerated Linear Algebra}
\newacro{HLV}[HLV]{Hanford, Livingston and Virgo}
\newacro{SNR}[SNR]{signal-to-noise ratio}
\newacro{IID}[IID]{independent and identically distributed}
\newacro{PSD}[PSD]{power spectral density}
\newacro{DAG}[DAG]{directed acyclic graph}

\newacro{CE}[CE]{common-envelope}
\newacro{OC}[OC]{open cluster}
\newacro{NSC}[NSC]{nuclear star cluster}
\newacro{CHE}[CHE]{chemically-homogeneous evolution}
\newacro{GC}[GC]{globular cluster}
\newacro{AGN}[AGN]{active galactic nuclei}
\newacro{SMT}[SMT]{stable mass transfer}

\setcounter{figure}{0}
\setcounter{table}{0}
\setcounter{equation}{0}

\renewcommand{\thefigure}{S\arabic{figure}}
\renewcommand{\thetable}{S\arabic{table}}
\renewcommand{\theequation}{S\arabic{equation}}

\newpage
\subsection{Supplementary Material}

\section{Model details}\label{app:model-details}

We consider two separate approaches to model how the effective spin distribution evolves with mass. The first one is the \Mixture model from Ref.~\citep{Alvarez-Lopez:inprep}, summarized in Eq.~\eqref{eq:mixture-def}. We use this model as a discovery engine to identify residual structure beyond the parametric model $p_\mathrm{par}(m_1,\chieff)=p_\mathrm{par}(m_1)p_\mathrm{par}(\chieff)$, where $p(m_1)$ is a mixture of two Gaussian peaks and a broken power-law tapered at low masses and $p(\chieff)$ is a \skewnormal distribution.

The flexible component, $p_\mathrm{PP}(m_1,\chieff)$, is built using two-dimensional \PixelPop with 100 bins in each dimension~\cite{Heinzel:2024jlc}, uniformly spaced in \chieff and $\ln m_1$ over $\chieff \in [-1,1]$ and $m_1 \in [3,300]\,\Msun$. \PixelPop is a high-resolution, non-parametric model that infers the joint distribution of (a subset of) \acs{BBH} parameters with minimal astrophysical assumptions. As in Refs.~\cite{Alvarez-Lopez:2025ltt,LIGOScientific:2026ctl}, we take the \ac{ICAR} limit to improve computational efficiency. However, whereas the original implementation of \PixelPop infers the merger rate in each bin, here we adapt \PixelPop to infer a normalized probability density. That way, we can mix it with the normalized parametric component via a mixture weight $\xi$. We then separately infer the overall merger-rate scale, $R_0$. In addition, rather than sampling the coupling parameter between bins, $\ln \sigma$, which controls the smoothness of the \ac{ICAR} field, we marginalize over it. Further details on the implementation of the \Mixture model are provided in Ref.~\citep{Alvarez-Lopez:inprep}.

In the \gtwou model, we write the conditional distribution of effective spin given primary mass as:
\begin{widetext}
\begin{equation}
\begin{split}
p(\chieff \mid m_1) &=
\xi_G(m_1)\,
\mathcal{G}_{\epsilon}\!\big(\chieff;\, \mu_{\chi_{\rm eff}},
\sigma_{\chi_{\rm eff}},\epsilon_{\chi_{\rm eff}}\big)\, + \\
&\big[1 - \xi_G(m_1)\big]
\Big\{
\xi_U^+(m_1)\, U^+(\chieff;\chi_{\rm eff}^{\max})
+ \big[1 - \xi_U^+(m_1)\big]\,
U^-(\chieff;\chi_{\rm eff}^{\min})
\Big\}.
\end{split}
\label{eq:gtwou}
\end{equation}
\end{widetext}

Here $\mathcal{G}_{\epsilon}$ is a \skewnormal distribution
truncated to $\chieff\in[-1,1]$, with location $\mu_{\rm eff}$,
scale $\sigma_{\rm eff}$, and skewness parameter
$\epsilon_{\rm eff}$~\citep{Banagiri:2025dxo}. The functions
$U^+$ and $U^-$ are half-uniform distributions on the positive and
negative sides of $\chieff=0$, respectively. Their inner edges are fixed at $\chieff=0$, while the outer edges, $\chi_{\rm eff}^{\max}$ and $\chi_{\rm eff}^{\min}$, are inferred from the data. The three components evolve with primary mass through mixing fractions $\xi_G(m_1), \xi_U^+(m_1) \in [0, 1]$: $\xi_G(m_1)$ controls the fraction of systems in the narrow, slowly spinning Gaussian bulk, with the remainder $1-\xi_G(m_1)$ assigned to the broader non-Gaussian component. Within this non-Gaussian component, $\xi_U^+(m_1)$ determines the fraction with positive $\chieff$, while $1-\xi_U^+(m_1)$ controls the fraction with negative $\chieff$. Further details about the mixing fractions --- specifically, how we assess whether a component is required by the data or not --- are given in the following section. We use the same $m_1$ parametrization as in the \Mixture model. We implement the \gtwou model using the \textsc{GWPopulation} package~\citep{Talbot:2024yqw,Talbot:2019}. We use the \textsc{Dynesty} nested sampler~\citep{Speagle_2020} to produce posterior samples and use 2000 live points in our analysis. 

For both the \Mixture and \gtwou models, the remaining binary parameters are modeled using one-dimensional independent models. We use a power-law evolution in redshift $z$~\citep{Fishbach:2018edt}, a power-law distribution for the mass ratio $q\in(0,1]$, and a skewed Gaussian for the effective precessing spin $\chip$~\citep{Racine:2008qv}, disjoint from \chieff. In the \Mixture model, we fix the minimum of the mass ratio distribution to $q_{\min}=0.1$, as in the \PixelPop analyses in Ref.~\citep{LIGOScientific:2026ctl}. The corresponding parameters, priors, and descriptions of our models are listed in Table~\ref{tab:model-params}. We note that, in the \gtwou model, we adopt a prior on $\sigma_{\chi_{\mathrm{eff}}}$ that is narrower than in the \Mixture model to mitigate label-swapping degeneracies between the Gaussian and uniform components.

We consider 259 events from \acs{GWTC-5} with false-alarm rates below $1\,\mathrm{yr}^{-1}$~\cite{LIGOScientific:2026wfs}, using the individual-event posterior samples publicly released by the \acs{LVK} collaboration~\citep{ligo_scientific_collaboration_and_virgo_2026_20348005,ligo_scientific_collaboration_and_virgo_2026_20348006}. We estimate the population likelihood using Monte Carlo integrals and limit the variance to be less than one during inference~\citep{Tiwari:2017ndi,Essick:2022ojx,Talbot:2023pex,Heinzel:2025ogf}. Our waveform choices largely follow Ref.~\citep{LIGOScientific:2026ctl}. However, we reduce the parameter estimation contribution to the log-likelihood variance by using the \textsc{IMRPhenomXPHM}~\citep{Pratten:2020ceb} samples for GW200129. Selection effects are computed using sensitivity estimates for \acs{BBH} systems across O1--O4b, obtained via a simulation campaign~\citep{Essick:2025zed,ligo_scientific_collaboration_2026_19500052}.

\section{Further details on the mixing fractions of the \gtwou model}

One of our main goals with the \gtwou model is to determine whether each component is required by the data as a function of primary mass. To do so, the model must allow individual components to vanish at a given mass, which occurs when the relevant mixing fraction is exactly 0 or 1. When $1-\xi_G(m_1)=0$, for instance, the $\chieff$ distribution is described entirely by the Gaussian component, whereas when $1-\xi_G(m_1)=1$, it is described entirely by the two half-uniform components. Similarly, within the non-Gaussian component, $\xi_U^+(m_1)=1$ corresponds to a purely positive half-uniform contribution, while $\xi_U^+(m_1)=0$ corresponds to a purely negative half-uniform contribution.

Previous approaches have modeled mass-dependent mixing fractions by defining a flexible spline in primary mass and applying a sigmoid transformation to restrict them to the interval $(0, 1)$~\citep[e.g.,][]{Plunkett:2026pxt}. However, a sigmoid only approaches the boundary values 0 and 1 in the limit $\mp\infty$. We therefore use a spike-and-slab prior~\citep{Mitchell:1988} on each mass-dependent mixing fraction, which preserves the flexibility of a spline in primary mass, while placing finite prior probability at the boundary values $0$ and $1$.

For each $\xi(m_1)$, we define a set of latent spline nodes $\{u_i\}$, placed uniformly in $\log m_1$ over $m_1\in[3,200]\,\Msun$. We use 10 nodes for $\xi_G$ and 8 nodes for $\xi_U^+$. The node values are assigned independent standard-normal priors, $u_i \sim \mathcal{N}(0,1)$, and are interpolated with a cubic spline to obtain the latent function $u(m_1)$. We then map $u(m_1)$ to the unit interval using
\begin{equation}
\xi(m_1) =
\begin{cases}
0 &
\Phi\!\left[u(m_1)\right] \leq \gamma \\[0.6em]
\dfrac{\Phi\!\left[u(m_1)\right]-\gamma}{1-2\gamma} &
\gamma < \Phi\!\left[u(m_1)\right] < 1-\gamma \\[1.0em]
1 &
\Phi\!\left[u(m_1)\right] \geq 1-\gamma,
\end{cases}
\label{eq:clip-prior}
\end{equation}
where $\Phi$ is the standard-normal cumulative distribution function and $\gamma=0.25$ controls how much prior mass is placed at the boundaries of the unit interval. Since $\Phi(u_i)$ is uniformly distributed on $[0,1]$, the transformation in Eq.~\eqref{eq:clip-prior} places 25\% of the prior mass at $\xi(m_1)=0$, 25\% at $\xi(m_1)=1$, and distributes the remaining 50\% uniformly across $(0,1)$. 

\begin{table}[t]
\centering
\begin{tabular}{lll}
\hline\hline
\multicolumn{1}{c}{\textbf{Symbol}} &
\multicolumn{1}{c}{\textbf{Description}} &
\multicolumn{1}{c}{\textbf{Prior}} \\
\hline
\multicolumn{3}{c}{\textbf{Common parameters}} \\
\hline
$\lambda$ & Redshift power-law index & $\mathcal{U}(-2,10)$ \\
$\mu_{\chi_p}$ & Mean of the $\chi_p$ Gaussian & $\mathcal{U}(0,1)$ \\
$\sigma_{\chi_p}$ & Width of the $\chi_p$ Gaussian & $\mathcal{U}(0.005,1)$ \\
$\epsilon_{\chi_p}$ & Skewness of the $\chi_p$ Gaussian & $\mathcal{U}(-1,1)$ \\
$\mu_{\chi_{\rm eff}}$ & Location of the $\chi_{\rm eff}$ Gaussian & $\mathcal{U}(-1,1)$ \\
$\epsilon_{\chi_{\rm eff}}$ & Skewness of the $\chi_{\rm eff}$ Gaussian & $\mathcal{U}(-1,1)$ \\
$\alpha_1$ & Power-law index below break & $\mathcal{U}(-4,12)$ \\
$\alpha_2$ & Power-law index above break & $\mathcal{U}(-4,12)$ \\
$m_\mathrm{break}$ & Power-law break location & $\mathcal{U}(20,50)$ \\
$\mu_1$ & Location of the first peak & $\mathcal{U}(5,20)$ \\
$\sigma_1$ & Width of the first peak & $\mathcal{U}(0,10)$ \\
$\mu_2$ & Location of the second peak & $\mathcal{U}(25,60)$ \\
$\sigma_2$ & Width of the second peak & $\mathcal{U}(0,10)$ \\
$\lambda_0$ & Power-law mixing fraction & Dirichlet \\
$\lambda_1$ & First peak mixing fraction & Dirichlet \\
$\lambda_2$ & Second peak mixing fraction & Dirichlet \\
$\delta_{\min,1}$ & Low-mass smoothing range in $m_1$ & $\mathcal{U}(0,10)$ \\
$m_{\min,1}$ & Minimum $m_1$ mass & $\mathcal{U}(3,10)$ \\
$\beta$ & Mass-ratio power-law index & $\mathcal{U}(-2,7)$ \\ 
\hline
\multicolumn{3}{c}{\textbf{\Mixture model parameters}} \\
\hline
$\xi$ & Mixing fraction; see
Eq.~\eqref{eq:mixture-def} & $\mathcal{LU}(10^{-5},1)$ \\
$q_{\min}$ & Minimum mass-ratio value  & $=0.1$ \\
$\sigma_{\chi_{\rm eff}}$ & Width of the $\chi_{\rm eff}$ Gaussian & $\mathcal{U}(0.005,1)$ \\
$m_{\max}$ & Maximum mass & $=300\,\Msun$ \\
\hline
\multicolumn{3}{c}{\textbf{\gtwou model parameters}} \\
\hline
$u_i$ & $i^\mathrm{th}$ spline-node latent value & $\mathcal{N}(0,1)$ \\
$\sigma_{\chi_{\rm eff}}$ & Width of the $\chi_{\rm eff}$ Gaussian & $\mathcal{U}(0.005,0.15)$ \\
$\chi_{\rm eff}^{\max}$ & Right edge of the $\chi_{\rm eff}>0$ uniform & $\mathcal{U}(0.05,1)$ \\
$\chi_{\rm eff}^{\min}$ & Left edge of the $\chi_{\rm eff}<0$ uniform & $\mathcal{U}(-1,-0.05)$ \\
$\delta_{\min,2}$ & Low-mass smoothing range in $m_2$ & $\mathcal{U}(0,10)$ \\ 
$m_{\min,2}$ & Minimum $m_2$ mass & $\mathcal{U}(3,m_{\min,1})$ \\
$m_{\max}$ & Maximum mass & $=200\,\Msun$ \\
\hline\hline
\end{tabular}
\caption{Priors on the model parameters, grouped into parameters common to both the \Mixture and \gtwou models (top), unique to the \Mixture model (middle), and unique to the \gtwou model (bottom). Here, $\mathcal{LU}$ corresponds to a log-uniform prior. }
\label{tab:model-params}
\end{table}

Because the prior assigns finite probability to both boundaries, we can directly test whether a component is required or not at any given primary mass. Each boundary value $b\in\{0,1\}$ of a mixing fraction removes one specific component from the model. For $1-\xi_G(m_1)=b$: $b=0$ removes the non-Gaussian component, and $b=1$ removes the Gaussian bulk. For $\xi_U^+(m_1)=b$: $b=0$ removes the positive uniform, and $b=1$ removes the negative uniform. We therefore consider two mutually exclusive hypotheses: $\mathcal{H}_{\rm abs}(b)$, in which $\xi(m_1)=b$ and the associated component is absent, and $\mathcal{H}_{\rm pres}(b)$, in which $\xi(m_1)\neq b$ and the component is present. We calculate the Bayes factor in favor of the presence of the component by dividing the posterior odds by the prior odds:

\begin{align}
\label{eq:component-bf}
\mathcal{B}(m_1;b)
&=
\frac{p[\mathcal{H}_{\rm pres}(b)\mid d]} {p[\mathcal{H}_{\rm abs}(b)\mid d]}  \frac{\pi[\mathcal{H}_{\rm abs}(b)]}{\pi[\mathcal{H}_{\rm pres}(b)]} \nonumber\\ 
&=
\frac{p[\xi(m_1)\neq b\mid d]}{p[\xi(m_1)=b\mid d]}
\frac{\pi[\xi(m_1)=b]}{\pi[\xi(m_1)\neq b]},
\end{align}
where $p$ and $\pi$ denote the posterior and prior probabilities, respectively. In other words, we reweight the two hypotheses to equal prior odds, which removes the approximate $(1-\gamma):\gamma=3:1$ prior preference for $\mathcal{H}_{\rm pres}(b)$, induced by our choice of $\gamma=0.25$. Throughout this \emph{Letter}, we report $\ln\mathcal{B}(m_1;b)$, so that $\ln\mathcal{B}(m_1;b)=0$ corresponds to $1:1$ odds between the two hypotheses, while positive and negative values favor the presence and absence of the component, respectively. In practice, we calculate Eq.~\eqref{eq:component-bf} over a mass interval $\Delta m_1$.

We also report the uncertainty due to finite sampling in $\ln\mathcal{B}$. We can rewrite Eq.~\eqref{eq:component-bf} using $p[\xi(m_1)\neq b\mid d] = 1 - p[\xi(m_1)= b\mid d]$ and $\pi[\xi(m_1)\neq b] = 1 - \pi[\xi(m_1)=b]$. We estimate each boundary probability as the fraction $f=N(b)/N$, where $N(b)$ is the number of samples satisfying $\xi(m_1)=b$ for all $m_1\in\Delta m_1$, and $N$ is the total number of samples in the corresponding posterior or prior set. The probability that the component is present is therefore $1-f$, and the corresponding log odds are $\ln[(1-f)/f]$. The fraction $f$ is a binomial estimator with variance $\sigma^2(f)=f(1-f)/N$; propagating to the log odds yields $\sigma^2\{\ln[(1-f)/f]\}=1/[Nf(1-f)]$. Since the posterior and prior samples are independent, we add their variances in quadrature to obtain $\sigma_{\ln\mathcal{B}} = \sqrt{\sigma_p^2+\sigma_\pi^2}$, where $\sigma_p$ and $\sigma_\pi$ are the uncertainties in the posterior and prior log odds, respectively. 

\section{Robustness to leave-one-out analysis for GW241011}\label{app:no-gw241011}
\begin{figure}
\centering
\includegraphics[width=\columnwidth]{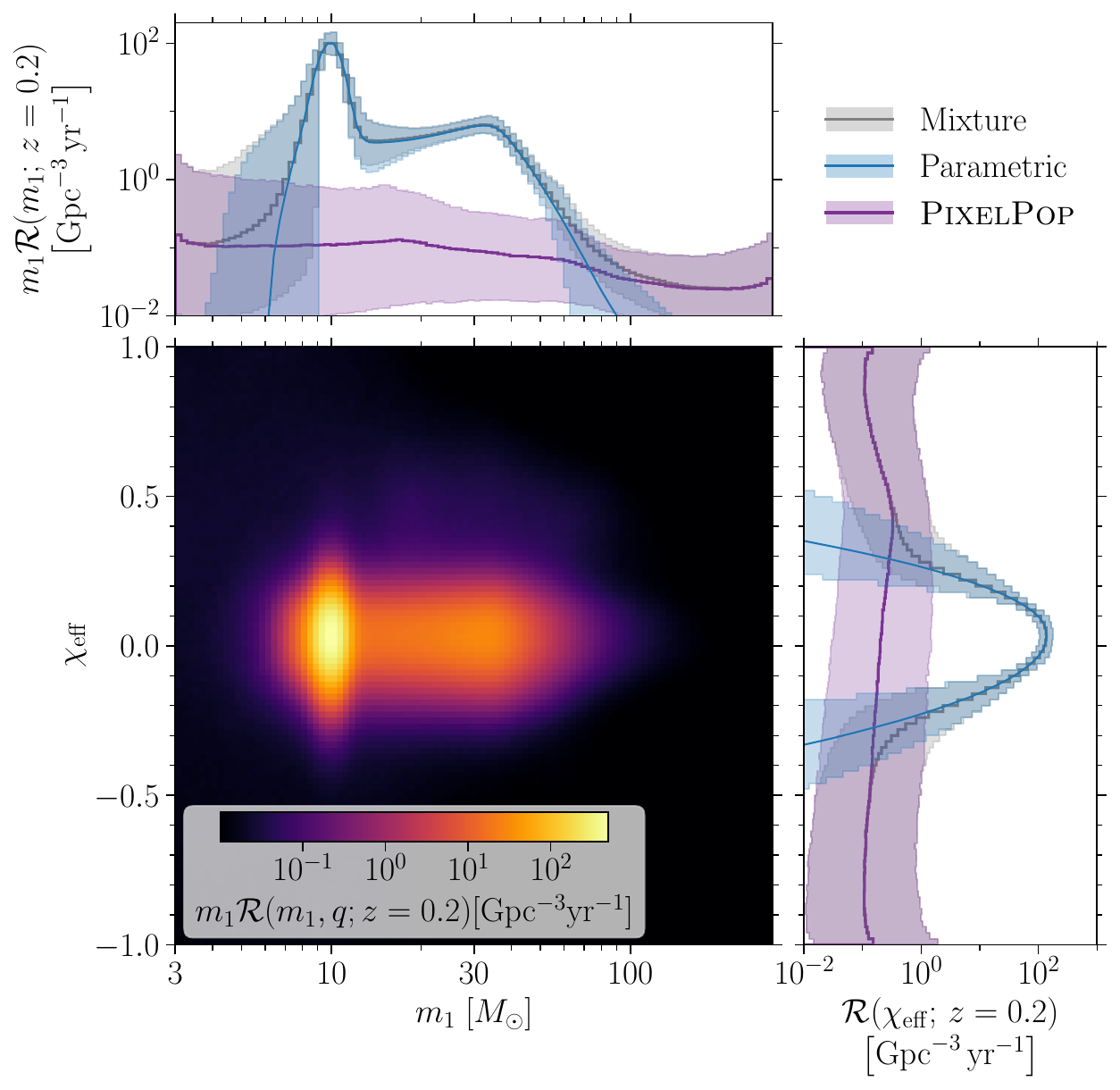}
\caption{Same as Fig.~\ref{fig:mixture-corner}, but removing GW241011 from the analysis.}
\label{fig:mixture-corner-no-GW241011}
\end{figure}

As a robustness check, we test whether the residual structure inferred by the \Mixture model is driven by GW241011~\cite{LIGOScientific:2025brd}, a highly spinning, unequal-mass \ac{BBH} system in \acs{GWTC-5}~\citep{LIGOScientific:2026ctl}, performing a leave-one-out analysis.

Figure~\ref{fig:mixture-corner-no-GW241011} shows the inferred comoving merger-rate density, $\mathcal{R}(m_1,\chieff;z=0.2)$, marginalized over mass ratio and effective precessing spin, for the catalog excluding GW241011. The results are consistent with those shown in Fig.~\ref{fig:mixture-corner}. The skewness of the parametric $\chieff$ component remains centered near zero, $\epsilon_{\chi_\mathrm{eff}} = -0.08_{-0.47}^{+0.58}$. The inferred \PixelPop mixing fraction is smaller in this case, $\xi = 1.6^{+5.6}_{-1.3}\%$, as expected after removing the event most strongly associated with the $\chieff$ residual (see the next section). However, the flexible \PixelPop component continues to identify residual structure in the $m_1$--$\chieff$ distribution. In the primary-mass range associated with GW241011, we still find a mild preference for positive effective spins within the \PixelPop component, with a 72\% probability that aligned-spin systems outnumber anti-aligned ones. These results indicate that the residual structure in \chieff is not driven by a single event but is a statistical property of the ensemble of \ac{BBH} mergers.

We also perform this leave-one-out analysis for the \gtwou model. In the mass range relevant for GW241011, $m_1\in[16,20]\,\Msun$, we find that the positive uniform component remains strongly required, with $\ln \mathcal{B}=6.6\pm 0.5$. Although this value is smaller than the corresponding Bayes factor reported in the main text ($\ln \mathcal{B}=8.4\pm 1.0$), the data still strongly disfavor a model without a positive-$\chieff$ contribution. In contrast, the preference for the negative uniform component is only mild, with $\ln \mathcal{B}=0.84\pm 0.03$ (c.f. $\ln \mathcal{B}=0.77\pm 0.03$). Thus, even after excluding GW241011, the data continue to support an excess at positive $\chieff$, which is difficult to explain by hierarchical mergers in dense stellar clusters alone.

\section{Individual-event Bayes factors} \hspace{0mm}
In the \Mixture model, the Bayes factor for each event $i$ to come from \PixelPop over the parametric \skewnormal component is~\cite{Plunkett:2026pxt}
\begin{table*}[t] 
\centering 
\begin{tabular}{lc|lc|lc}
\hline \hline
\textbf{Event} & 
$\boldsymbol{\ln\mathcal{B}_{\scriptstyle \mathrm{PP}/\mathrm{SN}}}$ & 
\textbf{Event} &  
$\boldsymbol{\ln\mathcal{B}_{\scriptstyle \mathrm{U}^{+}/\mathrm{SN}}}$  & \textbf{Event} & 
 $\boldsymbol{\ln\mathcal{B}_{\scriptstyle \mathrm{U}^{-}/\mathrm{SN}}}$
\\ \hline
GW241011 & 4.1 & GW241011 & 6.9 & GW191109 & 1.8 \\ 
GW241113 & 3.9 & GW241113 & 6.8 & GW241110 & 1.4 \\ 
GW231028 & 3.8 & GW231028 & 4.5 & GW241127 & 1.2 \\ 
GW231123 & 2.2 & GW231118\_005626 & 4.0 & GW241230\_233618 & 1.1 \\ 
GW190517 & 1.4 & GW190519 & 3.2 & GW241225\_082815 & 1.0 \\ 
GW231118\_005626 & 1.3 & GW190706 & 2.5 & & \\ 
GW240515 & 1.2 & GW230922\_040658 & 2.3 & & \\ 
GW190519 & 1.1 & GW190620 & 2.1 & & \\ 
GW190706 & 0.9 & GW231123 & 1.6 & & \\ 
GW190521 & 0.9 & GW230704\_212616 & 1.6 & & \\
 & & GW190517 & 1.4 & & \\
 & & GW240515 & 1.1 & & \\
 & & GW190602 & 1.0 & & \\
 & & GW170729 & 1.0 & & \\
\hline \hline
\end{tabular} 
\caption{Events with (natural) log Bayes factors higher than $0.9$ in favor of coming from the \PixelPop component rather than the \skewnormal component in the \Mixture model (\textit{left panel}); from the positive uniform component rather than the \skewnormal component in the \gtwou model (\textit{central panel}); from the negative uniform rather than the \skewnormal component in the \gtwou model (\textit{right panel}).} 
\label{tab:individual-events} 
\end{table*}

\begin{equation}\label{eq:BFs}
   \mathcal{B}_{\mathrm{PP/SN}} =\frac{ p(i \in \mathrm{PP}|\{d\})}{p(i\in \mathrm{SN}|\{d\})},
\end{equation}
where
\begin{equation}\label{eq:BF_integral}
\begin{split}
p(i \in \mathrm{PP}|\{d\}) =
\int d\Lambda\, d\theta_i\;&\xi\;
p(\Lambda\mid\{d\})\;p_{\rm PP}(\theta_i\mid\Lambda_{\rm PP}) \\
&\times \frac{p(\theta_i\mid d_i)}{p(\theta_i)}\, \frac{p(d_i)}{p(d_i\mid\Lambda)},
\end{split}
\end{equation}

and similarly for the \skewnormal component with $\xi \to 1-\xi$ and $p_{\rm PP} \to p_{\rm SN}$. The term $p(d_i|\Lambda)/p(d_i)$ can be written as $\int d\theta\, p(\theta|d_i) p(\theta|\Lambda)/p(\theta).$ With a similar procedure, we can calculate in the \gtwou model the Bayes factor for each event to come from component A instead of component B (with A, B being the \skewnormal, positive uniform, or negative uniform components). In practice, we compute these integrals as Monte-Carlo sums over individual-event samples $\theta_i$ and hyperposterior samples $\Lambda$.

\begin{figure}
\centering
\includegraphics[width=\columnwidth]{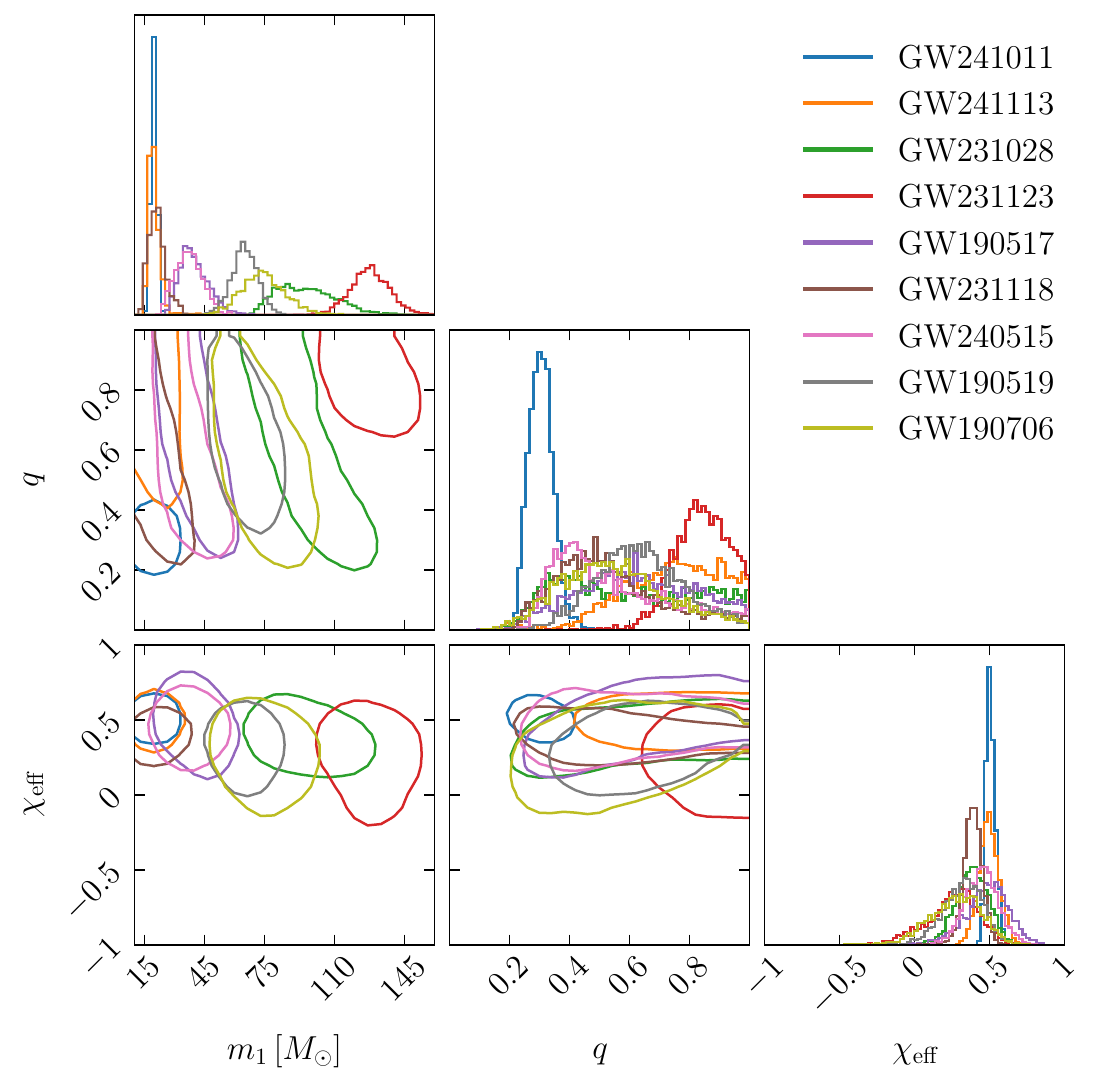}
\caption{Effective spin, mass ratio, and primary mass corner plot for the parameter-estimation samples for the events that appear in \emph{both} the left and central panel of Tab.~\ref{tab:individual-events}. The contours enclose 90\% credible regions. In the legend, GW231118 denotes the event GW231118\_005626.
}
\label{fig:event-corner-plots}
\end{figure}

Table~\ref{tab:individual-events} lists the events with the Bayes factor above $0.9$ in favor of assignment to the \PixelPop component in the \Mixture model (left panel) and to the positive and negative uniform components of the \gtwou model (central and right panel, respectively), relative to the \skewnormal component. Several of these events have also been identified by other analyses as candidates for hierarchical mergers~\citep{Tong:2025xir,Plunkett:2026pxt,Kimball:2020qyd}. Figure~\ref{fig:event-corner-plots} shows the individual-event
parameter-estimation samples for sources that appear in \emph{both} the left and central panel of Tab.~\ref{tab:individual-events}. These events all show significant support at positive \chieff; however, they span a wide range in primary mass and do not cluster at any particular mass ratio.

We also identify the events with the strongest preference for assignment to the negative uniform over the \skewnormal component in the \gtwou model. The leading event is GW191109, which ranks 11th in preference for the \PixelPop component in the \Mixture model. It is followed by GW241110, the exceptional event with strong support for a hierarchical origin~\citep{LIGOScientific:2025brd}. As discussed in the main text, however, the negative uniform is only mildly favored at the mass scale of GW241110, $m_1\in[16,20],\Msun$, with $\ln\mathcal{B}=0.77\pm0.03$. In the following section, we perform a leave-one-out analysis of the potentially contaminated event GW191109.

\section{Robustness to leave-one-out analysis for GW191109}

With $\chieff < 0$ at the 99.9\% credible level, GW191109 has the highest probability of assignment to the negative uniform component in the \gtwou model. However, it occurred near a glitch, and Ref.~\citep{Udall:2024ovp} found that its inferred \chieff depends strongly on the glitch-subtraction algorithm used. Given these data-quality concerns, we rerun the \gtwou model without GW191109 to test the robustness of the preference for a negative-\chieff component over $m_1 \in [47, 65]\,\Msun$. As discussed in the main text, this preference is weak even with GW191109 included: we obtain a mild $\ln{\mathcal{B}} = 2.6 \pm 0.1$ in favor of a negative-uniform component at this mass range.

\begin{figure}
\centering
\includegraphics[width=\columnwidth]{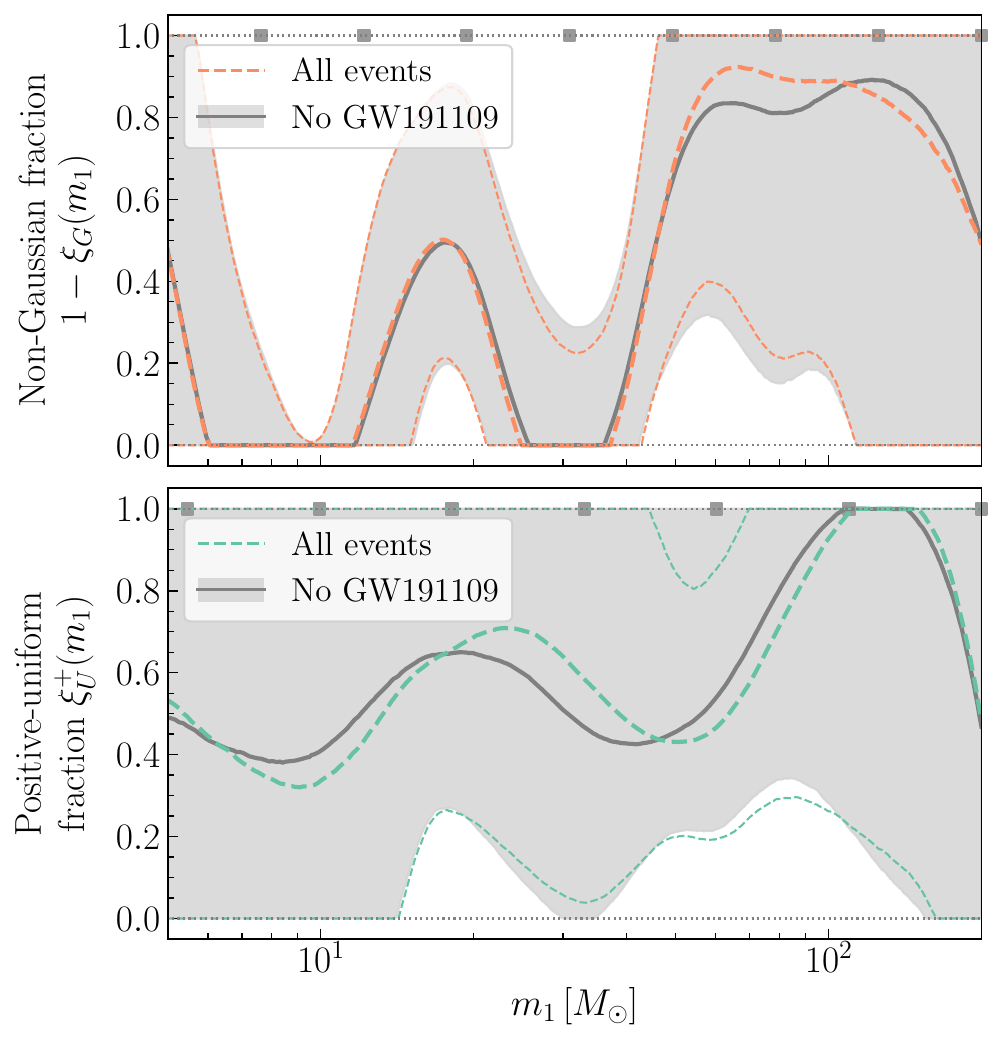}
\caption{Mixing fractions inferred with the \gtwou model as a function of primary mass. Dashed curves show the posterior median and bounds of the 90\% credible interval obtained using all events. Solid curves and shaded regions show the posterior median and 90\% credible interval obtained when excluding GW191109. All other features are as in Fig.~\ref{fig:mixing-fractions}.
}
\label{fig:mixing-fractions-noGW191109}
\end{figure}

Figure~\ref{fig:mixing-fractions-noGW191109} compares the 90\% credible regions for the mixing fractions inferred with and without GW191109. Although the non-Gaussian fraction does not change considerably, the positive-uniform fraction is now consistent with unity across the entire mass range: compared with the analysis that includes GW191109, the posterior probability that $\xi_U^+(m_1)<1$ decreases from 95\% to only 88\%. Consistently, we find that the preference for a model including the negative uniform reduces to $\ln\mathcal{B}=1.6\pm0.1$ (c.f. $\ln\mathcal{B}=2.6\pm0.1$). This reduction is equivalent to a decrease in the odds in favor of negative-\chieff sources---as expected from hierarchical mergers in dense stellar clusters---from $13\text{:}1$ to $5\text{:}1$ at this mass range.

\end{document}